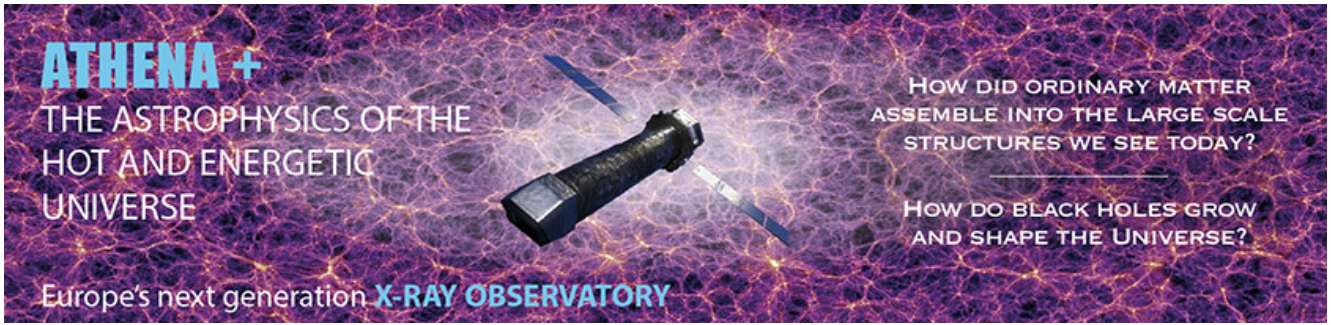

# The Hot and Energetic Universe

An *Athena+* supporting paper

## The X-ray Integral Field Unit (X-IFU) for Athena+

## Authors and contributors

**D. Barret, J. W. den Herder, L. Piro,** L. Ravera, R. Den Hartog, C. Macculi, X. Barcons, M. Page, S. Paltani, G. Rauw, J. Wilms, M. Ceballos, L. Duband, L. Gottardi, S. Lotti, J. de Plaa, E. Pointecouteau, C. Schmid, H. Akamatsu, D. Bagliani, S. Bandler, M. Barbera, P. Bastia, M. Biasotti, M. Branco, A. Camón, C. Cara, B. Cobo, L. Colasanti, J.L. Costa-Krämer, L. Corcione, W. Doriese, J.M. Duval, L. Fàbrega, F. Gatti, M. de Gerone, P. Guttridge, R. Kelley, C. Kilbourne, J. van der Kuur, T. Mineo, K. Mitsuda, L. Natalucci, T. Ohashi, Ph. Peille, E. Perinati, C. Pigot, G. Pizzigoni, C. Pobes, F. Porter, E. Renotte, J. L. Sauvageot, S. Sciortino, G. Torrioli, L. Valenziano, D. Willingale, C. de Vries, H. van Weers

## 1. EXECUTIVE SUMMARY

The *Athena+* mission concept is designed to implement the Hot and Energetic Universe science theme submitted to the European Space Agency in response to the call for White Papers for the definition of the L2 and L3 missions of its science program. The *Athena+* science payload consists of a large aperture high angular resolution X-ray optics and twelve meters away, two interchangeable focal plane instruments: the X-ray Integral Field Unit (X-IFU) and the Wide Field Imager (WFI). The X-IFU is a cryogenic X-ray spectrometer, based on a large array of Transition Edge Sensors (TES), offering 2.5 eV spectral resolution, with ~5" pixels, over a field of view of 5 arc minutes in diameter. In this paper, we briefly describe the *Athena+* mission concept and the X-IFU performance requirements. We then present the X-IFU detector and readout electronics principles, the current design of the focal plane assembly, the cooling chain and review the global architecture design. Finally, we describe the current performance estimates, in terms of effective area, particle background rejection, count rate capability and velocity measurements. Finally, we emphasize on the latest technology developments concerning TES array fabrication, spectral resolution and readout performance achieved to show that significant progresses are being accomplished towards the demanding X-IFU requirements.

## 2. THE ATHENA+ MISSION

Addressing the Hot and Energetic Universe science theme (Nandra, Barret et al. 2013, *Athena+ White Paper*) requires an X-ray observatory-class mission delivering a major leap forward in high-energy observational capabilities. Thanks to its revolutionary optics technology (Willingale, Pareschi et al. 2013, *Athena+ supporting paper*) and the most advanced X-ray instrumentation, the *Athena+* mission, will deliver superior wide field X-ray imaging, timing and imaging spectroscopy capabilities, far beyond those of any existing or approved future facilities. Like *XMM-Newton* today, *Athena+* will play a central role in all fields of astrophysical investigations in the next decade. No other observatory-class X-ray facility is programmed for that timeframe, and therefore *Athena+* will provide our only view of **the Hot and Energetic Universe**, leaving a major legacy for the future. The *Athena+* mission has an exceptionally mature heritage based on extensive studies and developments by ESA and the member states for *Athena*, IXO and XEUS. Compared with *Athena*, the *Athena+* concept incorporates important enhancements, including a doubling of the effective area (to 2 m² at 1 keV); an improvement in the angular resolution by a factor ~2 (to 5" on axis) and quadrupling of the fields of view of both the WFI and X-IFU, yet representing a realistic evolution in performance for a mission to fly in 2028. Table 1 summarizes the key mission requirements, as well as some comments on the enabling technology.

| Parameter | Requirements | Enabling technology/comments |
|---|---|---|
| **Effective Area** | 2 m² @ 1 keV (goal 2.5 m²)<br>0.25 m² @ 6 keV (goal 0.3 m²) | Silicon Pore Optics developed by ESA. Single telescope: 3 m outer diameter, 12 m fixed focal length. |
| **Angular Resolution** | 5" (goal 3") on-axis<br>10" at 25' radius | *Detailed analysis of error budget confirms that a performance of 5" HEW is feasible.* |
| **Energy Range** | 0.3-12 keV | Grazing incidence optics & detectors. |
| **Instrument Field of View** | *Wide-Field Imager:* (**WFI**): 40' (goal 50') | Large area DEPFET Active Pixel Sensors. |
| | *X-ray Integral Field Unit:* (**X-IFU**): 5' (goal 7') | Large array of multiplexed Transition Edge Sensors (TES) with 250 micron pixels. |
| **Spectral Resolution** | **WFI**: <150 eV @ 6 keV | Large area DEPFET Active Pixel Sensors. |
| | **X-IFU**: 2.5 eV @ 6 keV (goal 1.5 eV @ 1 keV) | *Inner array (10"x10") optimized for goal resolution at low energy (50 micron pixels).* |
| **Count Rate Capability** | > 1 Crab[1] (**WFI**) | *Central chip for high count rates without pile-up and with micro-second time resolution.* |
| | 1 mCrab, point source (X-IFU) with 90% of high-resolution events | *Filters and beam diffuser enable higher count rate capability with reduced spectral resolution.* |
| **Target of Opportunity Response** | 4 hours (goal 2 hours) for 50% of time | *Slew times <2 hours feasible; total response time dependent on ground system issues.* |

Table 1: Key parameters and requirements of the *Athena+* mission. The enabling technology is indicated.

---

[1] 1 Crab corresponds to a flux of ~2.4 10⁻⁸ ergs/s/cm² (2-10 keV).

As stated in the Hot and Energetic Universe white paper, the breadth of science affordable with the X-IFU touches to most of the key issues identified in the theme. For the hot Universe, this ranges from understanding how baryons accrete and evolve in the largest dark matter potential wells of groups and clusters (Pointecouteau, Reiprich et al., 2013, Ettori, Pratt et al. 2013, *Athena+ Supporting Papers*), how and when the energy contained in the hot intra-cluster medium was generated and how jets from AGN dissipate their mechanical energy in the intracluster medium (Croston, Sanders et al. 2013, *Athena+ Supporting Paper*). X-IFU will also find the missing baryons at *z*<2 and reveal the underlying mechanisms driving the distribution of this gas on various scales, from galaxies to galaxy clusters, as well as metal circulation and feedback processes (Kaastra, Finoguenov et al. 2013, *Athena+ Supporting Paper*). Similarly, for the energetic Universe, the X-IFU will enable the most obscured objects to be unveiled by revealing strong reflected iron lines. X-IFU observations will probe the first generation of stars to understand cosmic re-ionization, the formation of the first seed black holes, and the dissemination of the first metals (Aird, Comastri et al. 2013, Georgakakis, Carrera et al. 2013, *Athena+ Supporting Papers*). They will also reveal how accretion disks around black holes launch winds and outflows and how much energy they carry (Matt, Dovciak et al. 2013, *Athena+ Supporting Paper*). Similarly, AGN outflows and the processes by which the energy and metals are accelerated in galactic winds and deposited in the circum-galactic medium will be studied in details (Cappi, Done et al. 2013, *Athena+ Supporting Paper*). More generally, by providing spatially resolved high-resolution X-ray spectroscopy, the X-IFU instrument will enable new science to be performed for a wide range of objects, of great interest to the whole astronomical community, from planets, stars, supernova, binaries up to the most distant gamma-ray bursts (Branduardi-Raymont, Sciortino, et al. 2013, Sciortino, Rauw et al., 2013, Motch, Wilms, et al., 2013, Decourchelle, Costantini, et al. 2013, Jonker, O'Brien et al., 2013, *Athena+ Supporting Papers*).

The illustration of the power of spatially resolved high-resolution X-ray spectroscopy is provided in Figure 1 below, taken from Croston, Sanders et al. (2013, *Athena+ Supporting Paper*). This is the simulation of a 50 ks X-IFU exposure of the Perseus cluster. Such measurements will enable us to pinpoint the locations of jet energy dissipation, determine the total energy stored in bulk motions and weak shocks, and test models of AGN fuelling so as to determine how feedback regulates hot gas cooling.

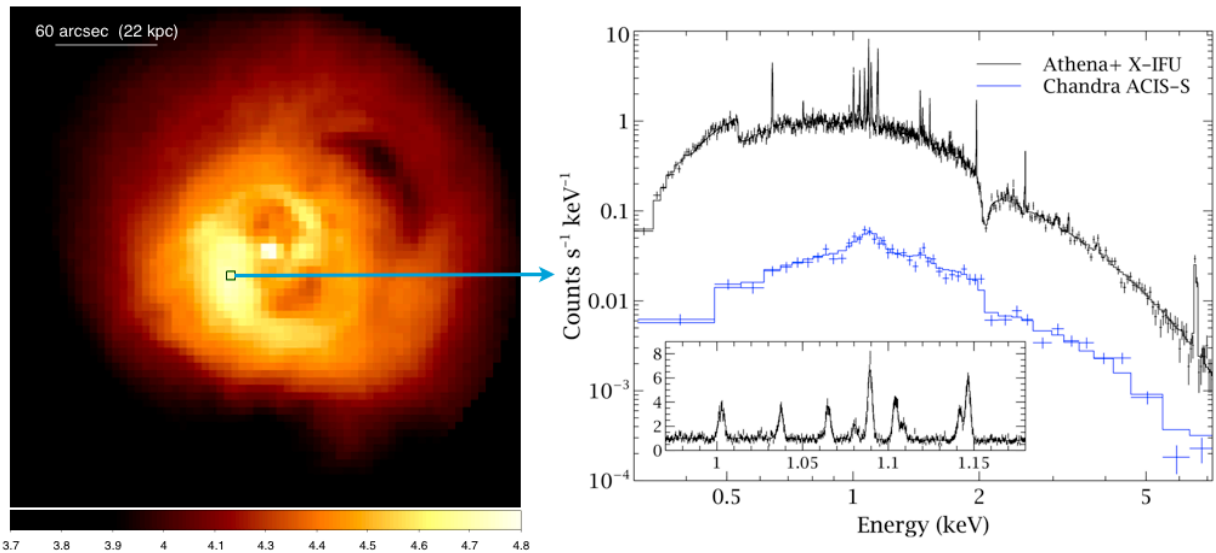

Figure 1: **Simulated *Athena+* observations of the Perseus cluster, highlighting the advanced capabilities for revealing the intricacies of the physical mechanisms at play.** The left panel shows a simulated 50 ks X-IFU observation (0.5-7 keV), displayed on a log scale. The spectrum on the right is from the single 5"×5" region marked by the box, with the existing *Chandra* ACIS spectrum for comparison. The inset shows the region around the iron L complex. With such observations velocity broadening is measured to 10-20 km s⁻¹, the temperature to 1.5% and the metallicity to 3% on scales <10 kpc in 20-30 nearby systems, and on <50 kpc scales in hundreds of clusters and groups. Such measurements will allow us to pinpoint the locations of jet energy dissipation, determine the total energy stored in bulk motions and weak shocks, and test models of AGN fuelling so as to determine how feedback regulates hot gas cooling (taken from Croston, Sanders, et al. 2013, *Athena+ Supporting Paper*).

## 3. THE X-IFU PERFORMANCE REQUIREMENTS

The X-IFU top-level performance requirements are listed in table 2. As described below, the requirements on the field of view and energy resolution can be achieved with an array of 3840 Transition Edge Sensors (TES) of 250 $\mu$m. A goal energy resolution of 1.5 eV could also be achieved using smaller pixel sizes (50 $\mu$m), but the details of the implementation of such a sensor remain to be defined.

| Parameter | Requirements |
|---|---|
| Energy range | 0.3-12 keV |
| Energy resolution: E < 7 keV | 2.5 eV (250 x 250 $\mu$m TES pixel) |
| Energy resolution: E > 7 keV | E/$\Delta$E = 2800 |
| Field of View | 5' (diameter) (3840 TES) |
| Detector quantum efficiency @ 1 keV | >60% |
| Detector quantum efficiency @ 7 keV | >70% |
| Gain error (RMS) | 0.4 eV |
| Count rate capability – faint source | 1 mCrab (>80% high-resolution events) |
| Count rate capability – bright source | 1 Crab (>30% low-resolution events) |
| Time resolution | 10 $\mu$s |
| Non X-ray background | < 5 10$^{-3}$ counts/s/cm$^2$/keV |

Table 2: Key performance requirements for the *Athena+* X-ray Integral Field Unit

## 4. X-IFU BASELINE DESIGN

This section provides the relevant information for the current X-IFU design, from the detection and readout principles to the mechanical, thermal and electrical architectures.

### 4.1. Detection principle

The TES micro-calorimeter senses the heat pulses generated by X-ray photons when they are absorbed and thermalized (see Figure 2). The temperature increases sharply with the incident photon energy and is measured by the change in the electrical resistance of the TES, which must be cooled at temperatures less than 100 mK and biased in its transition between super conducting and normal states (Irwin and Hilton, 2005, and references therein).

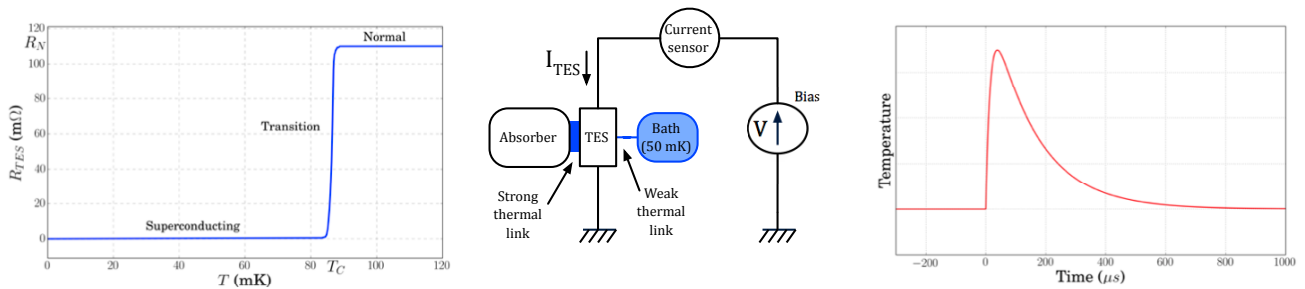

Figure 2: Principle of a TES (Transition Edge Sensor) acting as a micro-calorimeter. Left panel) The TES is cooled to lie in its transition between its superconducting and normal states. Middle panel) The absorption of an X-ray photon heats both the absorber and the TES through the strong thermal link. Right) The change in temperature (or resistance) with time shows a fast rise (due to the strong link between the absorber and the TES) and a slower decay (due to the combination of a weak link with the 50 mK thermal bath and a negative electrothermal feedback).

The absorber is 250 $\mu$m squared composed of 1 $\mu$m Au and 4 $\mu$m Bi to achieve the correct stopping power at 6 keV and low heat capacitance required for high energy resolution. We plan to use a Mo/Au bilayer TES for the X-IFU.

## 4.2.    SQUID readouts

The small bias current of the TES is readout using a low noise amplifier chain, consisting of a superconducting quantum interference device (SQUID) amplifier at 50 mK, followed by a semi-conductor low-noise amplifier (LNA). A SQUID amplifier is highly non-linear, showing a sinusoidal response. Its linearization is achieved with the use of a high gain feedback loop (see Figure 3).

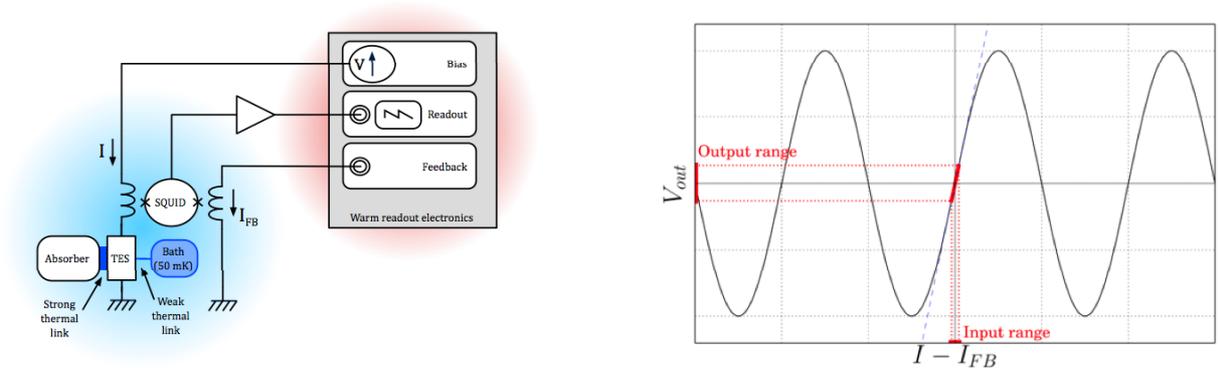

Figure 3: Left) SQUID readout of the TES with feedback. 2) Transfer function of the SQUID. Thanks to the feedback loop, the SQUID is restricted to operate in a linear regime.

## 4.3.    Multiplexed readout

Multiplexing enables several pixels to be readout by the same SQUID (Figure 4). For Frequency Domain Multiplexing (FDM), each pixel is AC biased with a specific carrier frequency. With a frequency range of ~1 to 5 MHz and a bandwidth separation of 100 kHz, 40 or more pixels can be multiplexed in a single read-out channel.

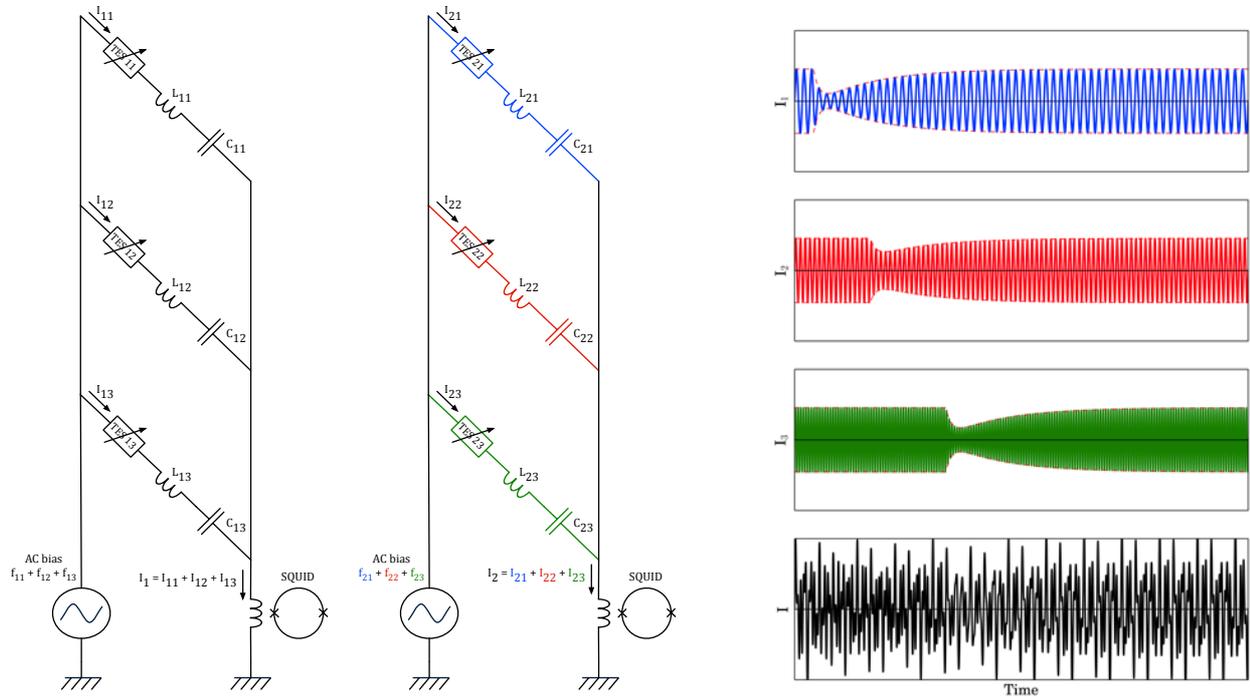

Figure 4: Frequency Domain Multiplexing principle: Left) The schematics of a channel of 3 pixels in 2 columns. Each pixel is biased at a specific frequency (f1, f2 and f3), each matching the resonant frequency of an $R_{TES}$ LC circuit. Right) For each pixel, the TES current modulated by the temperature/resistance variation induced by the absorption of an X-ray photon. The bottom panel shows the summed current readout by a single SQUID and the demodulated signal in the three pixels at the bottom.

FDM allows the reduction of the number of read-out channels and hence the thermal load on the detector. In our current baseline, **the 3840 TES sensors are readout in 96 channels of 40 pixels each.** A so-called base-band feedback (den Hartog et al. 2009, 2011) ensures that the feedback signal carrier is properly phased with the TES signal carrier at the SQUID input (propagation of the signal in the harness and digital electronics processing time introduces a phase delay that must be corrected for). De-modulation of the summed signal enables to reconstruct the shape of the signal in each pixel (see Figure 4, right panel). This signal is then sent to the event processor to determine the arrival time and energy of the incident X-ray photon.

## 4.4. Anticoincidence

Close underneath the TES array, an active anticoincidence detector must be placed to screen the particle background. This cryogenic anticoincidence detector is also based on TES technology, implementing a large area Silicon absorber with Iridium TES. This choice has the advantage, in addition to sharing several commonalities with the main TES array, of exploiting a fast component of TES (Figure 5), when working in the so-called a-thermal regime (Macculi et al. 2012, 2013). The energy released by the absorption of the photon in the absorber is rapidly converted into an a-thermal phonon population, not yet in thermal equilibrium with the system. These phonons can be then partly detected in the TES, giving rise to a fast contribution (green in Fig. 5) or thermalise in the absorber by inelastic scattering processes at the crystal surfaces leading to a temperature rise of the absorber and therefore delivering a slower, thermal contribution (blue) to the signal (Roth et al. 2008, Proebst et al. 1995). The cryogenic anticoincidence detector is constituted by a 4 TES-array and the related cryogenic SQUID and warm electronics. The main requirements, derived by imposing a rejection rate of the primary particles larger than 98%, are a low energy threshold of 20 keV, a size of 18 x 18 mm$^2$ (in 4 pixels, each 80 mm$^2$), a rise time less than 30 $\mu$s. Prototypes with performance close or within the requirements have been produced (see Macculi et al. 2012, 2013 and references therein).

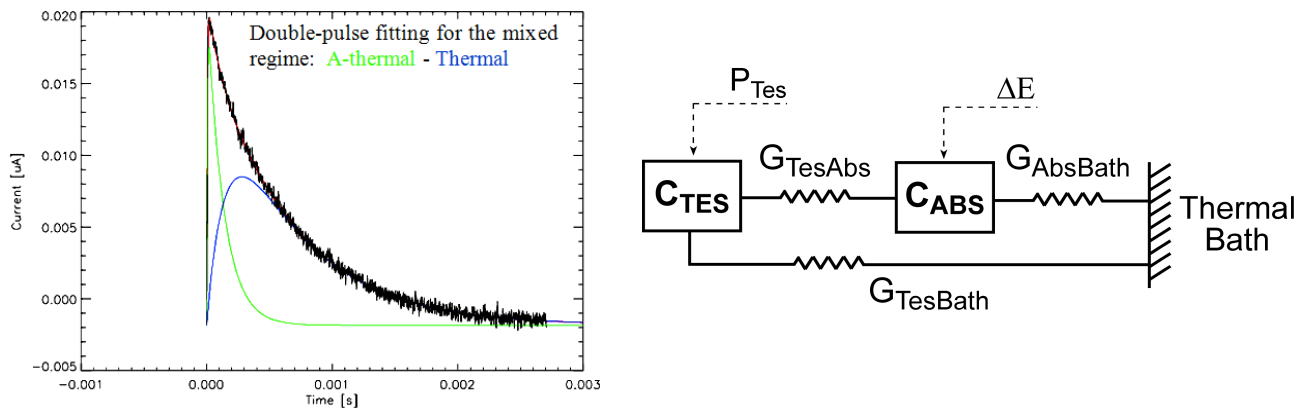

Figure 5: Left) The signal produced by the TES cryogenic anticoincidence showing a fast (1$\mu$s) rise, due to the a-thermal component (model in green) then followed by the slower thermal component (in blue). The thermal behavior is consistent with what expected by the thermal model of the system, shown on the right.

## 4.5. Focal plane assembly

The focal plane assembly provides the thermal and mechanical support to the sensor and the anti-coincidence detector. In addition it accommodates the cold electronics and provides the appropriate magnetic shielding. A magnetic field attenuation of 1.6x10$^5$ has been achieved by two shields: a super conducting Nb shield and a cryo-perm shield at 4 K and by an appropriate cooling sequence to avoid the trapping of magnetic flux. The focal plane assembly requires two optical blocking filters to reject the thermal and IR load on the detector (Figure 6). The current best estimate of the mass of the focal plane assembly is about 2 kg.

In addition, a filter wheel assembly will include filters to reduce the optical load on the detector for the observation of bright stars, a beam diffuser for observation of very bright (point) sources and a closed position to protect the detector against micrometeorites (a thick Be filter). The filter wheel will also include the calibration X-ray sources, needed to correct for slow (> 10 min) gain drifts in the detector.

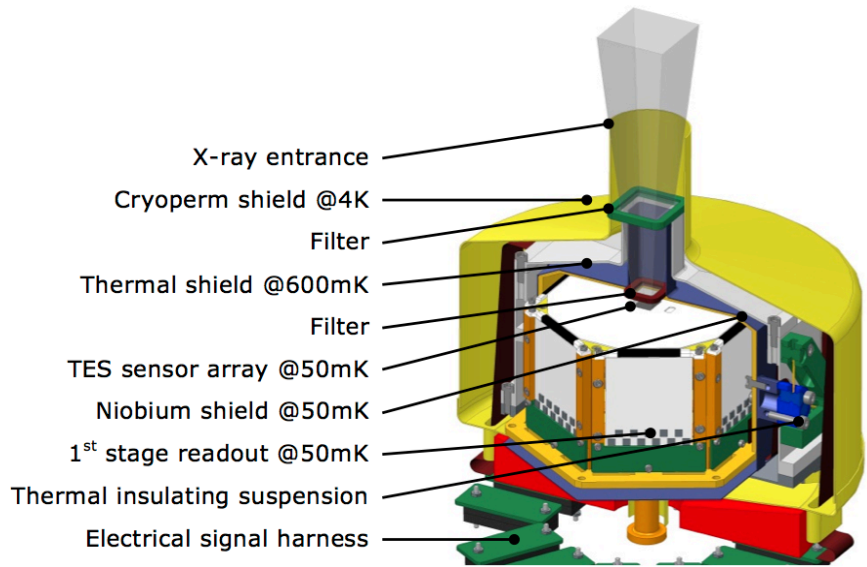

Figure 6: 3D drawing of the focal plane assembly. The outer shield is at 4 K. Between the detector, which is at 50 mK and the outer shield there is an intermediate radiation shield at 0.3 to 0.6 K (gray). The temperature stage is also used to thermally anchor the wiring. The electronics (SQUIDs, filters etc) is mounted on the side walls of the focal plane assembly to enable a compact design (white surfaces)

## 4.6.   Cooling chain

Thanks to many years of advanced cooler development within Europe, a high technology readiness level (TRL) for a fully European cooling chain is achievable by 2014-2015 (Geelen et al. 2006, Triqueneaux et al. 2008). Thus a single, technically sound, baseline has been proposed. Alongside the proposed baseline, different options in current development are available to serve as back-up coolers, in the event of late technology development, or if technical advantages are found in an alternative approach at a latter stage. Together with the cooler development shall also be developed in Europe the skills to design, build and integrate such a cryostat, via an ESA Core Technology Program in 2014. The cooling chain is a combination of several stages that can be divided into two major components: the upper cooling chain that provides cooling power at 2 K, and the last stage cooler which cools to 50 mK from this interface temperature (see Figure 7). The last stage cooler can be cycled at regular intervals whereas the upper cooling chain from room temperature provides continuous cooling (see Rando et al. 2010 for information about the IXO study).

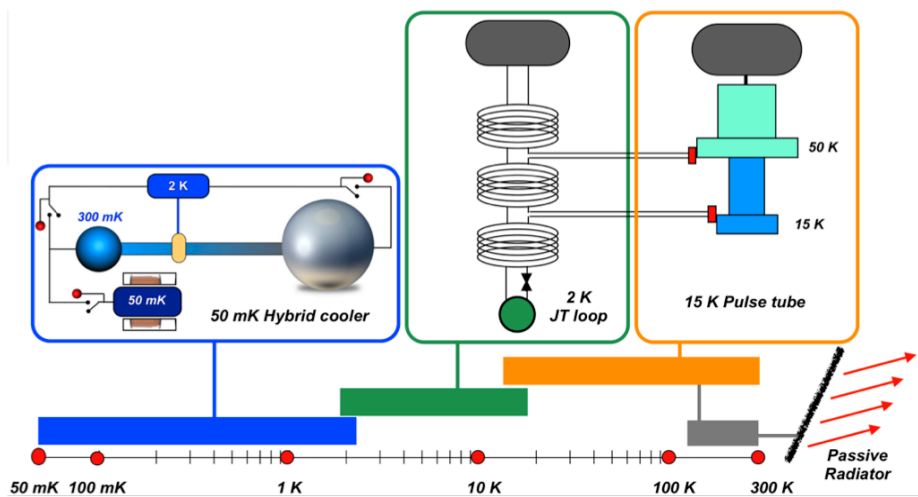

Figure 7: X-IFU proposed cooling chain comprises a combination of a small passive radiator, 15 K pulse tube coolers, 2K Joule Thomson loops and a 50/300 mK hybrid cooler.

The upper cooling chain comprises a small passive radiator, 15 K pulse tube coolers (Duval et al. 2013) and 2 K Joule Thomson loops currently under development at SFTC RAL. The latter two are currently in development and shall be qualified as part of ESA contracts.

As for the 50 mK last stage cooler, the development status is currently at an advanced stage (Duband et al. 2013). This hybrid cooler is based on the combination of a 300 mK sorption stage and a small adiabatic demagnetization stage.

A first engineering model has been developed and qualified in the framework of an ESA TRP focused on IXO (Figure 8). Duty cycles of 77% have been achieved for a hold time of 31 hours and 1 μW heat lift at 50 mK.

A second engineering model developed for the *Safari* instrument onboard *SPICA* will undergo a qualification program in late 2013. This particular unit has been designed to withstand static loads of 120 g and random vibration level of 21 g RMS weights 5 kg. It has been sized to provide net heat lifts of 0.4 and 14 μW respectively at 50 and 300 mK, for an overall cycle duration of 48 hours and a duty cycle objective of over 75%. For X-IFU the duty cycle should be significantly larger since the cooling power available at 2K is expected to be in the 20 mW range.

The full cooling chain, including the required number of redundant coolers, is expected to weight around 250 kg.

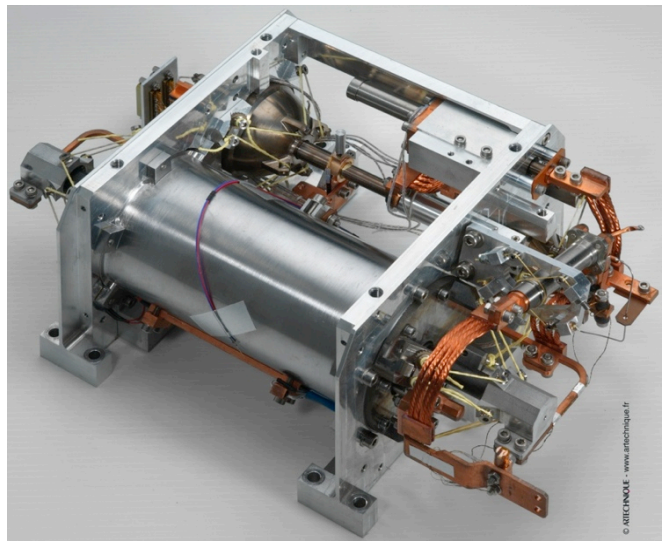

Figure 8: ESA TRP 50 mK Cooler (1 μW @ 50 mK and 10 μW @ 300 mK, 77% duty cycle efficiency)

## 4.7.    X-IFU architecture

The X-IFU block diagram is shown in Figure 9. The cold part contains the sensor arrays and its anticoincidence, the SQUIDs, while the warm part contains the digital electronics, the anti-coincidence electronics, the cooler drive electronics and the filter wheel electronics. The instrument control unit provides the interface between the various electronics and the spacecraft. As the detector is sensitive to EMC, a dedicated and properly filtered power supply unit must be used. The TES biasing, the SQUID multiplexer control, the data digitization and the generation of the feedback signals take place in the digital electronics, which also contains the event processor (Figure 9). The latter includes two major functions: event triggering and pulse height analysis. The handling of the anti-coincidence detector is completely separated from the main event chain. The number of charged particles is sufficiently small that processing of the anticoincidence data can be performed on the ground. The total mass and power for the electronic units is estimated to be about 60 kg and 300 W, respectively.

This brings the total mass of the instrument to about 400 kg. The X-IFU requires about 1 kW of power. The expected telemetry rate is 64 kbit/s (typical) with a maximum of 840 kbit/s (bright sources).

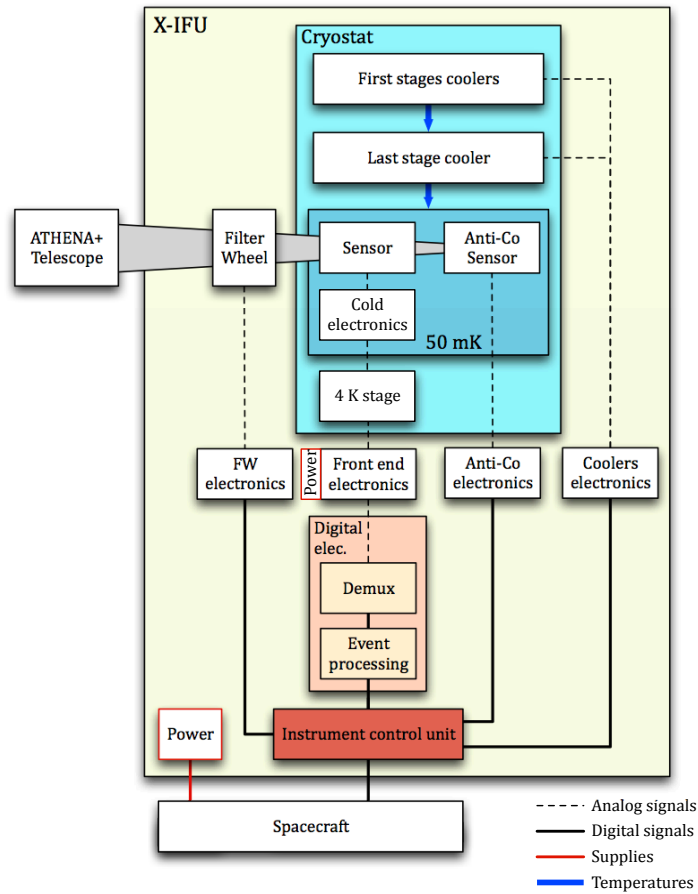

Figure 9: X-IFU block diagram. The TES array, its anticoincidence and the SQUIDs are at 50 mK. The front-end electronics consists primarily of the LNAs. The instrument control unit provides the interface with the spacecraft.

A close-up view of the full readout electronic chain is shown in Figure 10 below.

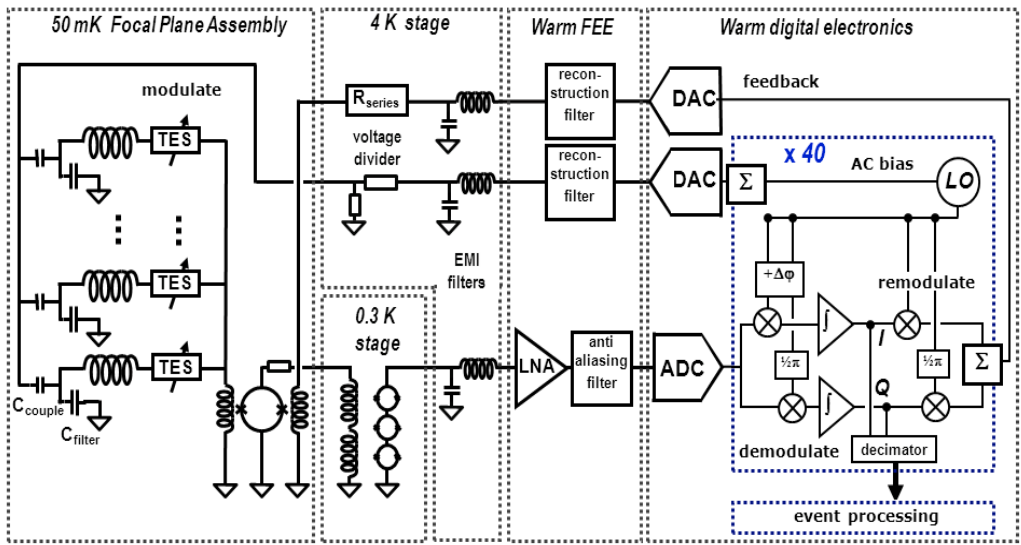

Figure 10: X-IFU close-up view of the front-end and digital electronics of the main TES array. The digital electronics is in charge of TES polarization, the SQUID multiplexer control, the data digitization and the generation of the feedback signals. The de-multiplexed signals are transmitted to the event processor. A simpler electronics is needed for the anticoincidence sensor (not shown here).

# 5. X-IFU PERFORMANCE ESTIMATES

In this section, we present our current best performance estimates of the X-IFU capabilities, matching the requirements of Table 2.

## 5.1. Effective area

The effective area of the instrument is a combination of the effective area of the optics, folded with the detector quantum efficiency, filling factor and the transmission of the fixed optical filters in front of the detector. These filters are required to allow for the cooling of the detector to 50 mK. The mirror effective area and the effective area including all detector related effects are shown in Figure 11. The X-IFU effective area around 1 keV is about 1.5 m² and is pretty flat across the X-IFU field of view.

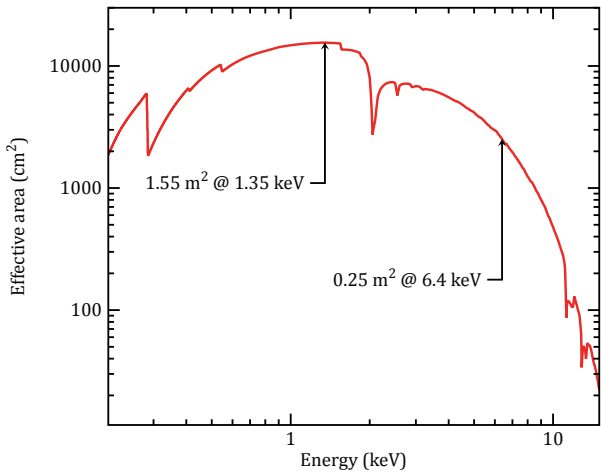

Figure 11: The X-IFU effective area, after accounting for all detector related effects (quantum efficiency, filters, filling factor, ...). The effective area of the X-IFU exceeds 1.5 m² at 1.35 keV and is 0.25 m² at 6.4 keV. Response matrix available at http://www.the-athena-x-ray-observatory.eu, under Resources.

## 5.2. Particle background rejection

The implementation of an anticoincidence system is required to meet the stringent scientific requirements on the instrumental background for the X-IFU operating in a L2 orbit (see Table 2). Cosmic ray and solar particles can interact either directly with the detector or via secondary particles produced in the material close to it. A fraction of these signals falls in the energy range of the detector and cannot be disentangled from true X-ray photons. The internal particle background has been estimated using GEANT4 simulations. This was necessary since no X-ray detector has ever flown so far in an L2 orbit. However, the particle background in this region is well characterized by several particle monitors flown on various satellites (e.g. Planck). The expected residual background without any shielding (Figure 12, left) would be 3 counts/cm²/s, ~60 times greater than the requirement.

With the implementation of the anticoincidence, the residual background decreases by a factor of about 10, essentially eliminating all primary particles. The residual non-rejected component is basically composed of secondary electrons. This component can be further reduced using a graded shielding with materials of a low yield for secondary electrons (Lotti et al. 2012a,b). The most suited material was found to be Kapton. This shielding allows us to meet the background requirements stated in Table 2.

In addition to the instrumental component, the background includes an X-ray diffuse component of various origins. A diffuse X-ray emission observed in every directions is produced at high energies mostly by the unresolved emission of AGNs, and below 1 keV by line emission from hot diffuse gas in the galactic halo and the local hot bubble, with contributions from Solar Wind Charge Exchange in the ¾ keV band. This latter component is highly variable, and a model representative of typical high galactic latitude fields had to be defined, as well as a study of its variations with time and pointing direction. Details on our modeling of these components are given in Lotti et al. (2013). The total background for X-IFU in the focal plane of *Athena+* is plotted in Figure 12 (right), showing that the particle background dominates only for energies above 2-3 keV.

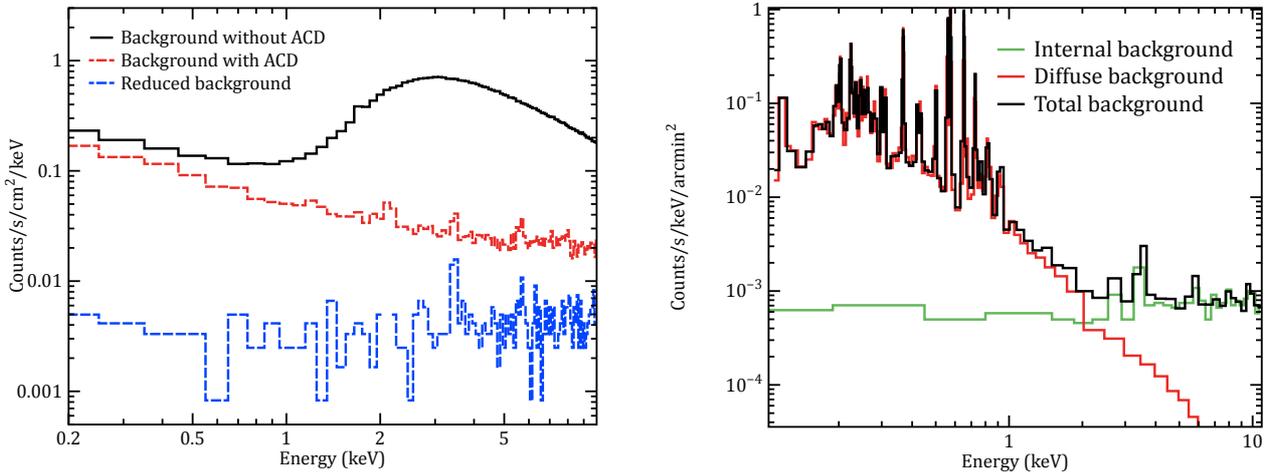

Figure 12: Left) The spectra of the internal background expected on X-IFU in all the cases analyzed. The black line is the expected residual background without any shielding, the red dashed line is the residual background with the implementation of the anticoincidence, and the blue dashed line is the final background level using a graded shield. Right) Background components for extended sources: the green line is the internal particle background, the red line is the diffuse component, and the total background is the black line.

## 5.3. Count rate capability

For a micro-calorimeter array the fraction of events with the highest spectral resolution depends on the rate of the incident photons (see den Herder et al. 2010, 2012). The rate in each individual pixel and therefore the fraction of high-resolution events for a given source is also determined by the shape of the mirror PSF and the relative pointing with respect to the pixels (the PSF can be centered at a pixel, just at the corner of four pixels or anywhere in between). In our simulations we assume that a period of $10\,\tau$ before and $40\,\tau$ after an event is necessary to achieve the maximum resolution, with a decay time of $\tau = 0.150$ ms (see Figure 13). Using appropriate filtering, the mid-res events typically have a resolution of about 3.5 eV which is still very good, while the low-res event resolution will be several tens of eV, depending on count rates. In Figure 13, the different event grades are shown as a function of total flux on the detector and in terms of the fraction of high-res, mid-res and low-res events. As can be seen, we can easily handle sources up to 1 mCrab without noticeable degradation of the resolution. At about 10 mCrab events are split more or less equally between the three different grades. In case of brighter sources one can insert a so-called beam diffuser in front of the detector. This is a small curved micro channel plate that diffuses the beam over a small circular ring on the detector (hence spreading the flux over more pixels). However, this approach has some disadvantages: the reflectivity of the beam diffuser is much less than 1 and the reflectivity is energy dependent.

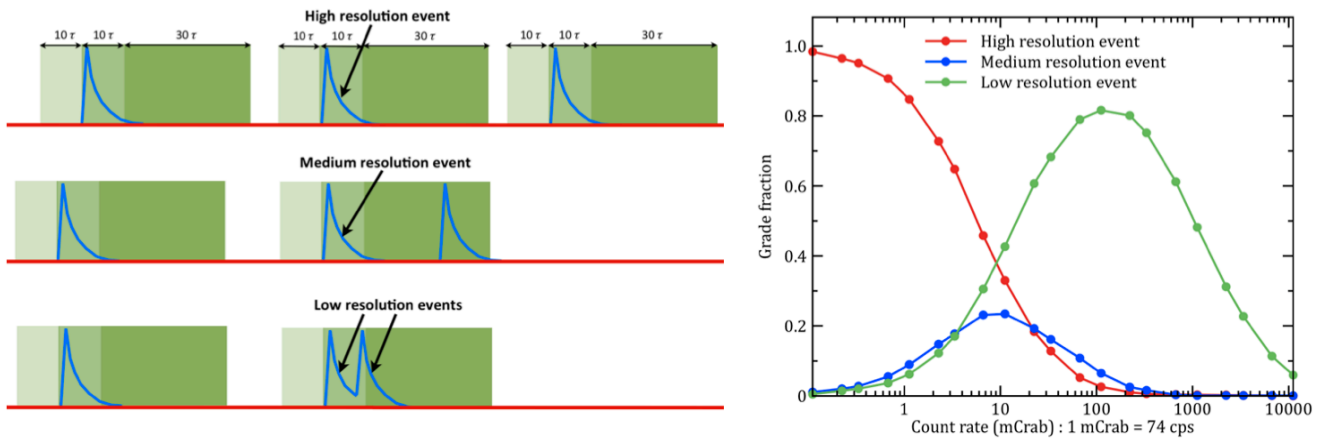

Figure 13: Left) The classification scheme of events. Right) X-IFU count-rate capability. The fractions of the different event grades are shown as a function of the total incident photon rate on the detector (red are high-res, blue: med-res, green: low-res).

### 5.4. Velocity measurements

One of the key science goals for the X-IFU instrument is the measurement of bulk and turbulent velocities in a variety of sources, such as clusters of galaxies. The accuracy with which these can be determined does not only depend on the statistics (the signal to noise ratio of the line) but also on the gain of the system and the knowledge about the shape of the redistribution function. Assuming that the uncertainty on the gain is 0.4 eV (rms), the minimum error on the bulk velocity is about 20 km/s. This is illustrated in Figure 14. For measuring turbulent velocities one needs to take into account the uncertainty in the known redistribution function. Assuming that the latter is also known to an accuracy of 0.4 eV (rms), one can compute the signal-to-noise ratio of the line required to ensure that the velocity is measured with an uncertainty of one fifth of its value. The minimum velocity broadening that can be measured by X-IFU corresponds to a velocity of about 30 km/s.

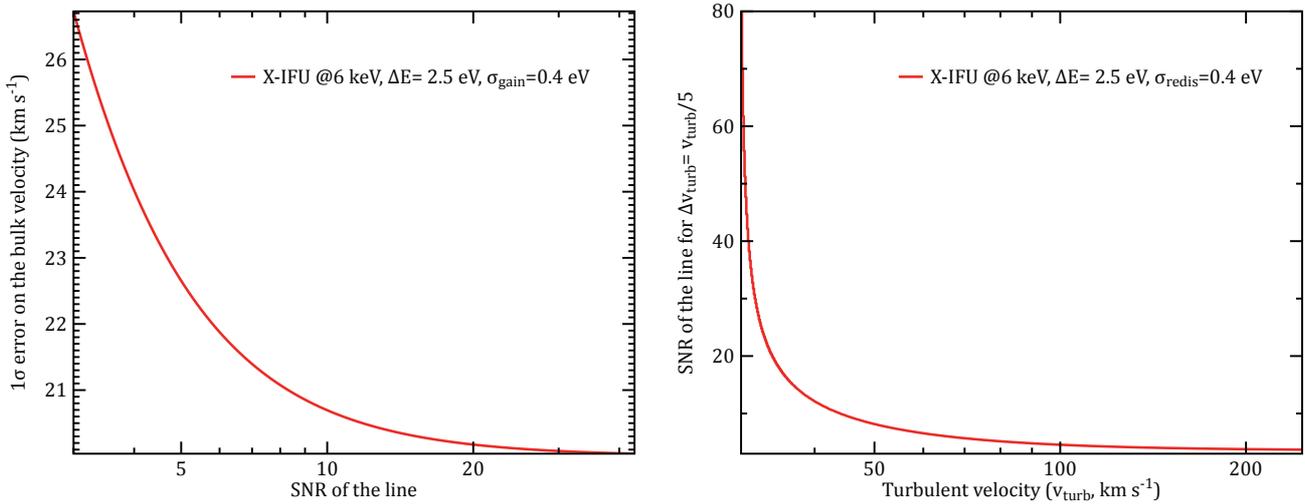

Figure 14: Left) 1 sigma error on the bulk velocity as a function of the signal-to-noise ratio of the line. This takes into account the statistical uncertainty and the instrumental uncertainty in locating the line (due to the gain variation), parameterized as a normal distribution with $\sigma_{gain}$=0.4 eV. Right) The signal-to-noise ratio of the line required to measure a turbulent velocity with an accuracy of 1/5 of the turbulent velocity (the minimum velocity broadening that can be measured with an accuracy of 20% is about 30 km/s, lower velocities are accessible but the uncertainty will increase).

## 6. X-IFU ON-GOING TECHNOLOGY DEVELOPMENTS

### 6.1. TES fabrication

In the development of arrays of microcalorimeters (TES + absorber) a number of specific requirements must be met, beside those pertaining to energy resolution and thermal time constant. In particular, keeping the thermal cross-talk between pixels as low as possible, providing a proper connection to the heat bath for each pixel, and achieving sufficient mechanical robustness of the detector, are relevant. Figure 15 shows the conceptual design of the microcalorimeter array pixels, together with pictures of its actual realization. The Ti/Au TES is designed such that it has a critical temperature of ~100 mK (note that for the X-IFU we will use Mo/Au bilayer TES). At the center of the TES a Cu/Bi absorber is lithographed using the recipe shown in Figure 16. The overhanging Cu/Bi absorber allows for close packing, with a 95% filling factor, a high absorption efficiency without a correspondingly high heat capacity, and it creates space for slots in the $Si_xN_y$ membrane to tune the heat conductance, and space for the connecting wiring. The design is realized with the so-called bulk micromachining route: deep vertical slots are etched in the backside of a Si (110) wafer by KOH etching. The advantage of bulk micromachining is its relative simplicity. To improve the cooling of the beams, 70% of their area is now covered with a 0.7 $\mu$m thick Cu coating applied by shadow deposition, which leads to excellent thermal properties. The TES and its absorber are positioned on a slotted membrane, which separate the membranes from each other and prevent thermal crosstalk between pixels. The width of the $Si_xN_y$-legs between the membranes and the Si beam is used to tune the thermal coupling of the pixel to the heat bath.

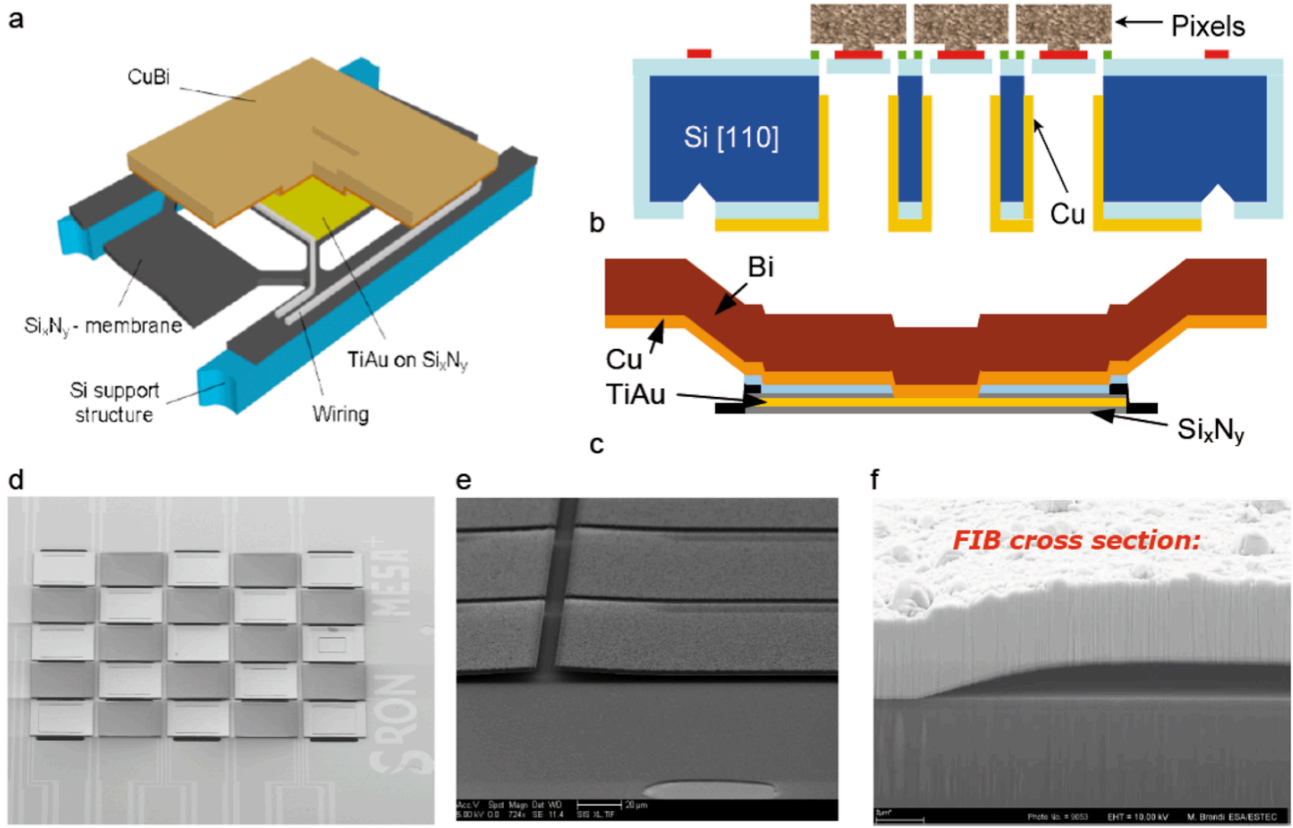

Figure 15: a. Exploded view of the pixel design with main materials indicated. b. Crosscut through support design. c. Crosscut through pixel design. d. SEM image of a 5x5 test array. e. Detail of the absorber structure. f. FIB cross-section of the joint between the absorber and the TES.

The details of this layout are continuously being improved. In particular on the following topics, investigated either within the Safari or the *Athena+* efforts, progress is planned:

The electrical wiring connection and current routing, in relation to the self-magnetic field and the weak-link effects, which will bring improvement in the energy resolution.

The connection between the absorber and the TES (stems). In recent years several groups have realized that smaller and properly placed absorber stems allows for a better energy resolution.

The method for fabrication of the Si support structure. Replacing the current KOH etching process with the Deep RIE (Bosch) process, will allow a better thermal conductance in lateral direction, and a greater mechanical rigidity, enabling us to fabricate arrays larger than the demonstrated arrays with 32 x 32 pixels.

The ultimate goal is to have a Mo/Au TES. In collaboration with several groups of CSIC (Spain), microcalorimeters based on Mo/Au TES's are being developed as an alternative to the Ti/Au route. The Mo/Au process is more challenging, but will bring the advantage of a higher stability of $T_C$ when the TES is exposed to high temperatures (i.e., no degradation has been seen after heating up the TES to 150 ºC).

Following a different approach, CEA/Saclay is developing high impedance Si:P:B thermometers, with superconducting composite tantalum absorbers, together with its cryogenic readout electronics (Sauvageot et al. 2013).

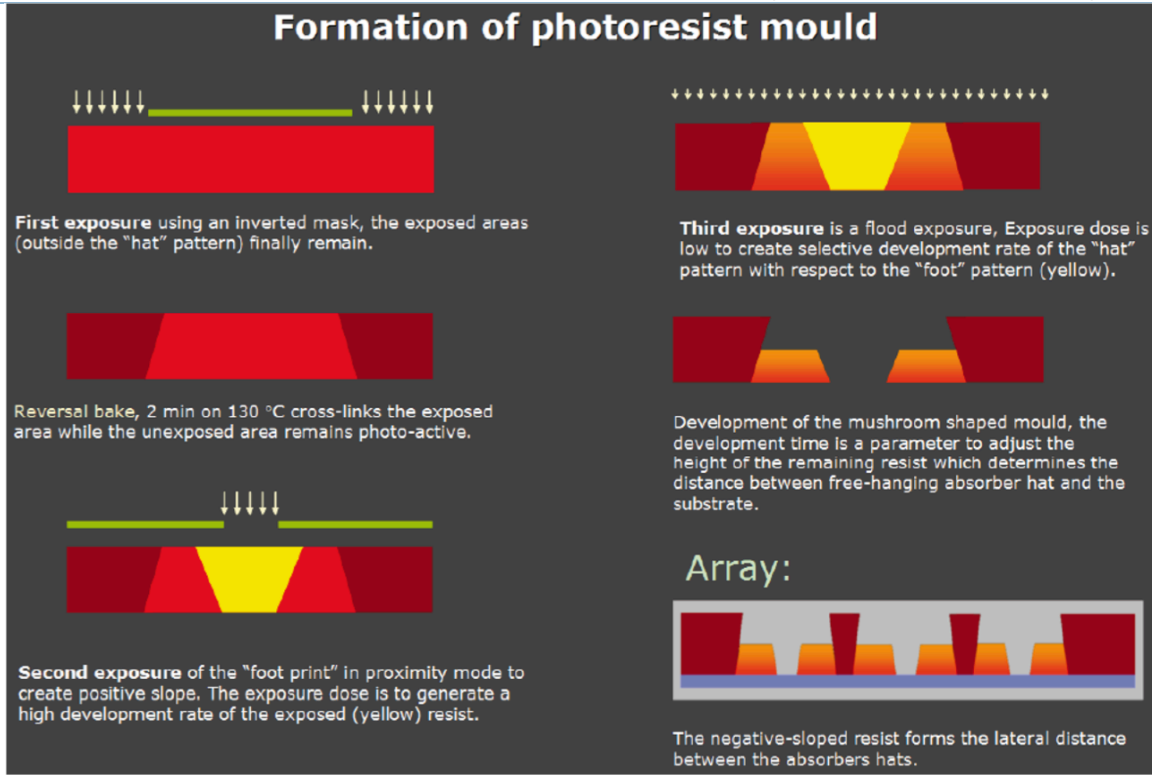

Figure 16: Details of the fabrication process for the mushroom mould with which the free-hanging CuBi absorber can be placed on top of the TES.

## 6.2.   Spectral resolution

The theoretical or intrinsic spectral resolution of a TES depends primarily on the heat capacitance of the absorber ($C$), the critical temperature (T), the resistance (R) and the sensitivity of the thermistor ($\alpha$) (e.g. Irwin and Hilton, 2005).

$$\Delta E \approx 2.35 \frac{1}{\sqrt{\alpha}} \sqrt{k_B T^2 C}, \alpha = T/R \times dR/dT \qquad (1)$$

Better energy resolution implies lower $T$ (hence a cryogenic detector), and/or a larger transition steepness (narrower transition, Figure 2) and/or lower heat capacitance. In practice, however, the spectral resolution will degrade when including the readout noise, the electrical and thermal cross talks between pixels, the effects of micro-phonics on the detector and the effect of thermal load from the environment on the sensor.

The baseline resolution is a measurement of the energy resolution achievable by the detector and is defined as the integrated Noise Equivalent Power (NEP) given by:

$$\Delta E = 2.35 \times \left( \int_0^\infty \frac{4df}{NEP^2} \right)^{-1/2} \qquad (2)$$

An X-ray energy resolution of 2.3 eV has been demonstrated when the pixels are DC biased (Gottardi, et al. 2012). The most recent results obtained with Frequency Domain read-out are shown in Figure 17. On the left side the X-ray spectrum of a 1.3 MHz AC biased pixel is shown. The best single pixel X-ray energy resolution measured so far was 3.6 eV (Akamatsu, H., et al. 2012). This resolution was consistent with the baseline resolution calculated from the detector noise and it was limited by the SQUID read-out noise. On the right sight of Figure 17, we show the NEP of a pixel biased at 2.5 MHz with an improved read-out scheme. For comparison, the NEP spectra of the DC biased and the 1.3 MHz ac biased pixels are also shown in the plot. The baseline resolution calculated from the NEP spectra of Figure 17 corresponding to 2.5 MHz is 2.7 eV.

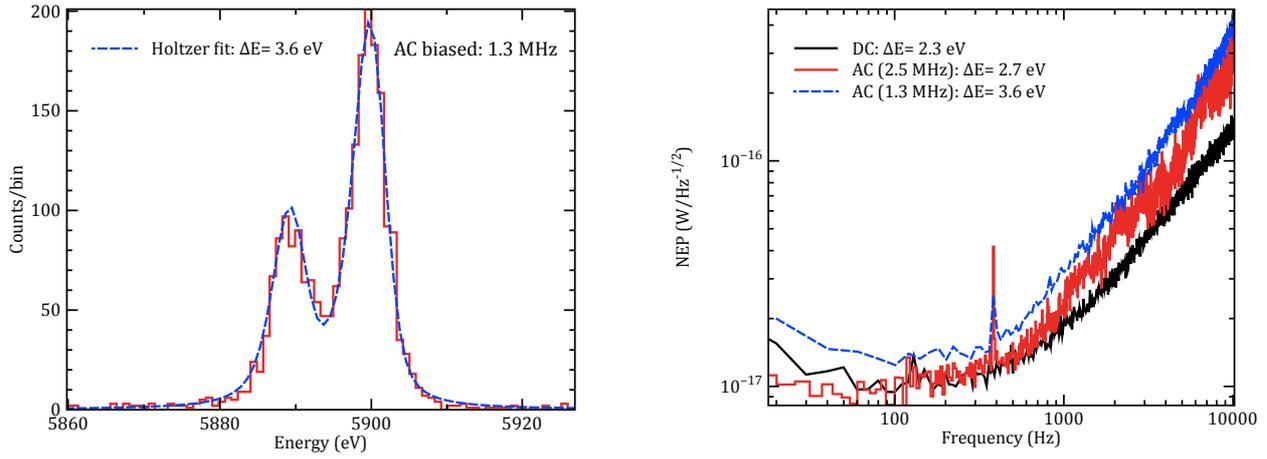

Figure 17) Left: measured X-ray resolution (Fe[55]) for a pixel biased under AC (1.3 MHz). Right: Noise Equivalent Power (NEP) for an AC bias pixel at 2.5 MHz with improved read-out (Red curve). Estimated baseline resolution is 2.7 eV. For comparison, the NEP spectra of the 1.3 MHz biased pixel ($\Delta$E=3.6 eV) and the DC biased pixel ($\Delta$E=2.3 eV) are shown by the blue and black lines respectively.

## 6.3.    Readout electronics

Many of the stringent requirements on the performance of the read-out system have already been demonstrated. This is in a large part owing to the synergy with the readout system of the far-IR *Safari* instrument for the JAXA mission *SPICA*, which allowed development to keep a constant pace. Although the TES sensors for *Athena+*, X-ray microcalorimeters, are quite different from the sensors that will be used in *Safari* far-IR bolometers, the readout systems are quite comparable, and so are the requirements pertaining to noise floor and crosstalk. Several requirements are even more stringent for *Safari*, in particular those for LC filter quality and frequency spacing and baseline stability.

Within the *Safari* development effort most of the relevant requirements, noise floor, LC filter performance, SQUID + LNA performance, dynamic range, gain bandwidth, have already been verified (den Hartog et al. 2012a,b). It was also demonstrated (for normal-driven TES sensors, i.e. resistors, which are easier to operate) that at least 56 channels could be multiplexed without degradation of the noise (den Hartog et al. 2012a), and that for TES far-IR bolometers the same low dark NEP levels (~0.45 aW/√Hz) can be obtained under DC and AC bias (Gottardi et al. 2012). 16-channel multiplexed read-out of X-ray detectors has been demonstrated in den Hartog et al. (2009). Figure 18 shows an example of the time sequences obtained in this experiment. The baseline resolution in this experiment was still 5 - 6 eV (calibrated to a 6 keV signal), due to the high noise floor of 40 pA/√Hz in this setup.

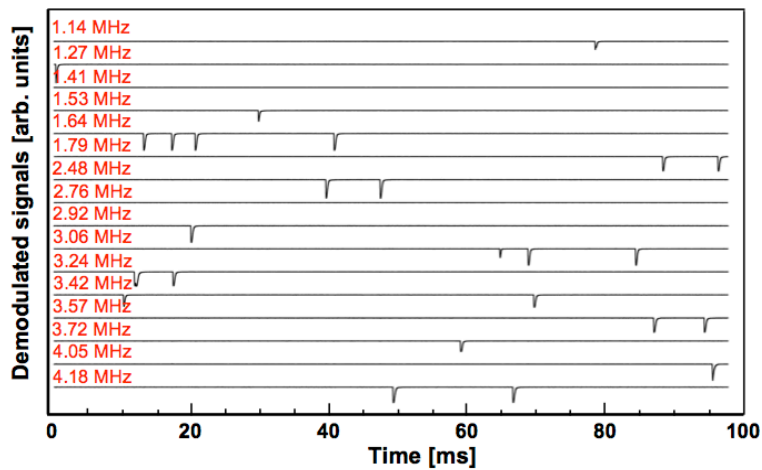

Figure 18: Time sequences a 16-channel multiplexed readout of an X-ray microcalorimeter array. Pulses in the different channels are clearly seen.

In parallel, noise floors at a 4 times lower level have been routinely shown (den Hartog et al. 2012a,b), which should limit the readout noise to 1.5 eV. In Figure 19 below we demonstrate the simultaneous multiplexed read-out of 38 TES bolometers (the far-IR sensitive variety of the TES sensors) with moderately low dark NEP levels (~1.3 aW/√Hz). This experiment shows that stable low-noise baselines are possible, that the fabrication of essential components such as LC filters is continuously improving, and that multiplexed readout does not degrade noise or detector performance, at least, at a ~1.3 aW/√Hz level for relatively slow bolometers. The 1/f features have a powerlaw of 1/3, appear to vary from run to run, are absent for pixels with an NEP equivalent to the TES X-ray sensors, and when they appear, the 1/f knee is well below frequencies relevant for the readout of fast X-ray pulses.

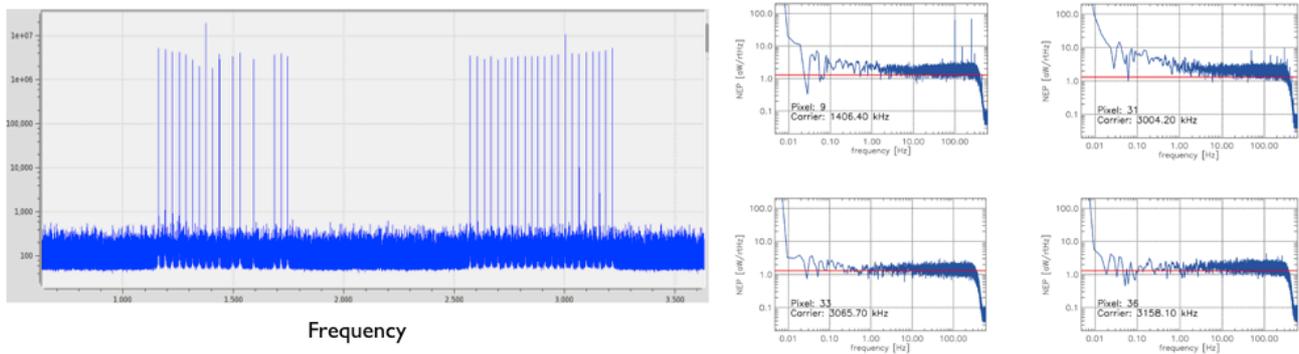

Figure 19: Left) Spectrum of the base-band feedback signal with 38 TES bolometers in closed loop. Right) Dark NEP curves which are the result of 400 seconds of multiplexed read-out of 38 TES bolometers. The horizontal line corresponds to the 1.3 aW/√Hz level.

## 7. CONCLUSIONS

The X-IFU instrument will provide unprecedented capabilities to perform spatially resolved high-resolution X-ray spectroscopy with arcsecond imaging. In view of the ambitious performance to be achieved, the X-IFU consortium is getting organized, while actively pursuing an aggressive development plan on the most critical technologies (cooling chain, sensors, readout electronics, ...). A demonstration model of the X-IFU around 2018 is foreseen. Very significant progresses have been achieved in the last few years, benefiting from earlier studies and an-going developments for *Safari/SPICA*. A powerful X-IFU instrument could then be built in Europe for a launch in 2028, under the leadership of France, The Netherlands and Italy. Nevertheless, even within a Europe led *Athena+* mission, foreign contributions (e.g. from Japan, United States, ...) potentially enhancing the instrument (e.g. in terms of design maturity) could also be considered.